# Maxwell's True Current

*Version 4*

Robert S. Eisenberg

*January 2, 2024*

*This is a reworked version of the preprint arXiv 2309.05667.*

Department of Applied Mathematics, Illinois Institute of Technology, Chicago IL 60616 USA

Department of Physiology & Biophysics; Rush University Medical Center; Chicago IL 60612 USA

Correspondence: bob.eisenberg@gmail.com

Telephone: +01-708-932-2597

C:\Users\bobei\Dropbox\2023 Maxwell's True Current, Revisited\arXiv Maxwell Current January 2024\Maxwell's True Current January 2, 2024.docx

**Abstract**

Maxwell defined a 'true' or 'total' current in a way not widely used today. He said that "... true electric current ... is not the same thing as the current of conduction but that the time-variation of the electric displacement must be taken into account in estimating the total movement of electricity". We show that the true or total current is a universal property of electrodynamics independent of the properties of matter. We use mathematics without the approximation of a dielectric constant. The resulting Maxwell Current Law is a generalization of the Kirchhoff Law of Current used in circuit analysis, that also includes the displacement current. The generalization is not a long-time low frequency approximation in contrast to the traditional presentation of Kirchhoff's Law.

The Maxwell Current Law does not require currents to be confined to circuits. It has been applied to three dimensional systems like the signaling system of nerve cells and muscle fibers. The Maxwell Current Law clarifies the flow of electrons, protons, and ions in mitochondria that generate ATP, the molecule that stores chemical energy throughout life. The currents are globally coupled because mitochondria are short. The Maxwell Current Law approach reinterprets the classical chemiosmotic hypothesis of ATP production. The conduction current of protons in mitochondria is driven by the protonmotive force including its component electrical potential, just as in the classical chemiosmotic hypothesis. The electrical potential is now the electrical potential as defined in the physical sciences by the Maxwell differential equations. The conduction current is now just a part of the true current analyzed by Maxwell. Details of the accumulation of charges do not have to be considered in the analysis of true current because true current does not accumulate. It is the true total current that provides the energy that generates the ATP, not just the protonmotive force.

Most scientists today are unaware of Maxwell's definition (Maxwell 1865b) of the 'true' current needed to estimate "the total movement of electricity":

> *"One of the chief peculiarities of this treatise is the doctrine which it asserts, that the true electric current, that on which the electromagnetic phenomena depend, is not the same thing as the current of conduction, but that the time-variation of the electric displacement, must be considered in estimating the total movement of electricity, so that we must write, [the sum] .... as an equation of true currents."*

Note that *'peculiarities'* is a Victorian word (in formal 19th Century English) equivalent to *'characteristics'* or *'features'* in modern scientific American/English.

> The text is from Vol. 2, Section 610, p. 232. The *'equation of true currents'* is in, eq H, Vol.2, Sect 610, p. 233 of (Maxwell 1865b). Maxwell uses the name *'Total current'* throughout his analysis 327-341 of Vol 1. This paper uses the names interchangeably, as defined in eq. (2).

**Maxwell's definition is esthetically appealing** because of its author, and its generality. It involves only two parameters, the electric and magnetic constants. These constants are universal and do not depend on the properties of matter, at all.

Maxwell's definition is helpful in numerical procedures (Qiao et al. 2023b, 2023a). Maxwell's definition applies to systems that involve three dimensions, like nerve fibers, and mitochondria of biological cells. Maxwell's definition is not confined to circuits although it (see eq. (2) and (5)) has been used to design circuits important throughout our technology since the invention of the telegraph.

It is interesting to compare Maxwell's idea of true current with the idea of current used in modern science. Many workers in chemistry and biology and engineering define current as conduction current and do not discuss 'true' current explicitly, with little emphasis on displacement current. They certainly do not take the approach *"that the time-variation of the electric displacement, must be considered in estimating the total movement of electricity"*. The total movement of electricity is often computed using only the flux of electrons, without explicit treatment of displacement current, and without time derivatives at all, in contradiction to Maxwell's words: the total current *"must be considered in estimating the total movement of electricity"*. In particular, I have not found a treatment of Kirchhoff's law that uses true current, or includes displacement current in its definition of current besides that of our group, e.g., (Eisenberg 2019c; Eisenberg et al. 2018).

Engineers include the displacement current when they analyze real circuits but not as Maxwell recommends. Engineers modify the circuit they analyze so it includes displacement current. Maxwell includes displacement current in the definition of true current.

Displacement current, and time derivatives, do not appear in traditional presentations of Kirchhoff's law. Examples of analysis of purely resistive circuits using Kirchhoff's law, do not include time derivatives in any treatment I know of. When they are actually constructed circuits containing only resistors—devoid of capacitors—have time dependence in their response to time dependent applied potentials (or currents). The time response arises from displacement currents that are an unavoidable property of systems and circuits, as Maxwell shows so clearly Vol. 2, Section 610, p. 232 of (Maxwell 1865b). A classical Kirchhoff analysis does not include these time dependent responses because it does not include time dependence in its definition of current.

Engineers are forced to include such displacement currents in their analysis if the results are to fit experiments on actual circuits. Engineers often introduce the displacement current by changing the circuit they analyze. They introduce supplementary circuit elements, called 'stray capacitances', instead of using true (total) current in Kirchhoff's law. These are discussed at length in this paper with mathematical precision, I hope, in eq. (7)-.(9)

Engineers know that without these fictitious elements their circuits do not function as they should. The stray capacitances describe displacement currents that must be included if circuits are to behave properly.(Horowitz and Hill 2015), p. 579-589. Feynman discusses this practice in some detail (Feynman, Leighton, and Sands 1963a), Vol 2, Section 22-23 although he too presents Kirchhoff's law as a low frequency approximation and uses currents without derivatives in his examples.

This paper looks for a way to use Maxwell's idea of true current in a modern context particularly useful in biology. We show that the idea of true current casts a useful light on the role of sodium and potassium fluxes in the action potential of nerve and muscle cells (Hodgkin and Rushton 1946; Davis and de No 1947; Huxley 1972).

We show that the idea of true current is useful in understanding the flow of electrons, ions, and protons in mitochondria. The chemiosmotic flow of electrons and protons (Morelli et al. 2019; Juhaszova et al. 2022; Boyer 1988; Wikstrom et al. 2015; Mitchell and Moyle 1967; Mitchell 1975, 1977; Stryer 1995; Mitchell 1961) in mitochondria has been a central topic in biology since the 1960's, because that flow creates ATP (Mitchell and Moyle 1967; Mitchell 1975, 1977; Berg, Tymoczko, and Stryer 2010; Nicholls 2013; Ogawa and Lee 1984; Walker 1998; Mitchell 1961). ATP stores chemical energy throughout living systems. The hydrolysis of ATP provides most of the energy of life.

We show mathematically that the idea of total or true current is needed if we want to estimate the '*total movement of electricity*' in mitochondria, just as Maxwell said. In mitochondria, the total movement of electricity involves sodium $Na^+$ and potassium $K^+$ ions, hydrogen ions, and, of course, electrons. We show in particular that the movements of protons across the membrane of mitochondria are coupled to the movements of all charges by the current laws eq. (2) & (5). It is the true total current that provides the energy that generates the ATP, not just the protonmotive force.

It is easy for physicists and engineers to be confused by the use of the word 'proton' in biology.(DeCoursey 2023; DeCoursey and Hosler 2014; Chaves et al. 2023) The word 'proton' in the relevant biological literature is shorthand for positively charged water, which can take on many chemical forms (Yamashita and Voth 2012; Knight and Voth 2011; Knight et al. 2010; Chen et al. 2007), depending on the chemical and biological context, for example, depending on the structure of the proteins it moves through. Proton movement is not controlled by just the protonmotive force of the original chemiosmotic hypothesis. Proton movement is a form of electrical current and so it follows the same physical laws as any other current. These are the Maxwell equations of electrodynamics that imply—by mathematics alone without approximation or additional physics—the laws of total current flow eq. (2) and (5). Current laws couple the protonmotive force to the electric field, and thus to the movement of all charges across membranes, as well as displacement current across the membrane capacitance. This coupling is not discussed in the classical literature of mitochondria, or the chemiosmotic hypothesis, as far as I know. Treating the flow of electrons (and ions and protons) in mitochondria the way they are treated in physical sciences seems to be helpful, as well as necessary (in the mathematical sense) in understanding mitochondrial function.

**Fundamental Equations of Electricity.** The partial differential equations introduced by Maxwell, summarized in (Maxwell 1865a), have been used for more than a century to describe the properties of electricity and its dynamics, in the form given in modern textbooks, for example, (Griffiths 2017), (Purcell and Morin 2013), (Zangwill 2013), and (Feynman, Leighton, and Sands 1963a). These hardly differ from Maxwell's original formulation, although the ordering is different, with Maxwell choosing to make the Maxwell-Ampere circuit equation his equation **A** perhaps because the circuit equation explains how light is an electromagnetic phenomena that can propagate through a near vacuum devoid of charges.

We start with the Maxwell-Ampere partial differential equation the circuit equation that describes how the magnetic field **B** is created by changes in the electric field **E** and the conduction current **J**. The charge and current are parsed in an unconventional manner here for reasons discussed below, immediately after variables are defined. The Maxwell-Ampere equation is written here without the dielectric constant $\varepsilon_r$ for reasons discussed at length previously (Eisenberg 2019b; Eisenberg 2019a; Eisenberg 2020b, 2021) $\varepsilon_r$ is more formally called the 'relative permittivity'. Dielectric currents are included in **J**.

**Ampere-Maxwell Equation** $\quad \frac{1}{\mu_0} \operatorname{curl} \mathbf{B} = \mathbf{J}_{true} = \mathbf{J} + \varepsilon_0 \, \partial \mathbf{E}/\partial t \quad$ **J** is zero in a vacuum $\quad$ (1)

$\mathbf{J}_{true}$ is Maxwell's 'true current' (defined above in words) displayed prominently as his eq. **A**, p. 465, 480 of (Maxwell 1865a). Perhaps Maxwell identified this equation as his first, his **A** equation because of the importance of displacement current in the propagation of light from the sun through the near vacuum of space. $\varepsilon_0$ is the electric constant. $\mu_0$ is the magnetic constant. $\mathbf{J}_{true}$ is **not** zero in a vacuum.

The parsing used here allows the fundamental eq. (1) to be independent of all other parameters. No properties of matter are included explicitly in this version of the Ampere-Maxwell equation (1). The equation applies, as is, in the vacuum of space and everywhere else. The properties of matter are described separately in equations that are solved together with eq.(1).

**Electro-mechanical model of polarization.** The Ampere-Maxwell Equation (1) needs to be supplemented by an electromechanical model of polarization. The electromechanical responses of matter to changes in the electric field is diverse, to understate things (Eisenberg 2019b; Eisenberg 2019a; Eisenberg 2020b, 2021), and so the electromechanical model of polarization is a phenomenological description of how the charge density in matter varies with field strength (etc.) as discussed later. Its role is similar to the role of density variation (compressibility) in fluid mechanics of simple or complex fluids.

The conduction current of any charge $\rho$ with mass is described by **J**, however brief, small, or transient is the current (or its flux). The conduction current **J** does not include the displacement current $\varepsilon_0\, \partial \mathbf{E}/\partial t$ found everywhere without exception. The conduction current includes all the movements of charge that occur when the electric field changes. Many of those movements of charge are nonideal and cannot be not well described by a single dielectric constant. (Eisenberg 2013; Barthel, Buchner, and Münsterer 1995; Barthel, Krienke, and Kunz 1998; Kremer and Schönhals 2003; Barsoukov and Macdonald 2018; Eisenberg 2019b; Eisenberg, Oriols, and Ferry 2017; Eisenberg, Oriols, and Ferry 2022) The conduction current also includes the ideal movements of charge that occur when the electric field changes. The conduction current **J** does include the idealized dielectric polarization current and the dielectric current $\varepsilon_0(\varepsilon_r - 1)$ of an ideal capacitor used in classical formulations and approximations of material systems. (Purcell and Morin 2013), p. 500-507 In a vacuum, the only current $\mathbf{J}_{true}$ is the displacement current $\varepsilon_0\, \partial \mathbf{E}/\partial t$ but that is not zero when (for example) light propagates from sun to earth. That vacuum displacement current $\varepsilon_0\, \partial \mathbf{E}/\partial t$ allows radiation like light to propagate in the space between the earth and the sun, which is nearly a vacuum.

Note that Maxwell's original statement of displacement current differs from that used here. Maxwell's original treatment included two components of displacement current. One of Maxwell's components was the displacement current of our eq. (1) $\varepsilon_0\, \partial \mathbf{E}/\partial t$ that is universal, present in a vacuum, and in matter, everywhere in fact, and has no adjustable parameters. The other component that Maxwell used $(\varepsilon_r - 1)\varepsilon_0\, \partial \mathbf{E}/\partial t$ depends on properties of materials. The material displacement current uses adjustable parameters and models to describe how material charge changes when the electric field changes. Maxwell's original treatment of displacement current can be reinstated from our equations in two steps. First, (1) by replacing $\varepsilon_0\, \partial \mathbf{E}/\partial t$ in our equations with $\varepsilon_r \varepsilon_0\, \partial \mathbf{E}/\partial t$ and then (2) by changing the definition of charge and conduction current correspondingly so displacement currents are not counted twice. These changes are easy to implement and mean that classical analysis can be easily modified if applications require it.

**Universal Law.** Current is parsed this way here so the Ampere-Maxwell equation (1) can be written as a universal law without explicit reference to the properties of matter. Without this rewriting, the fundamental field equations become constitutive, not universal laws, with the properties of matter—in all its diversity, with all its difficulties of measurement—embedded in its very definition of variables. The dependence of the classical Maxwell equations on approximate models of polarization is a problem that is widely recognized in texts by distinguished physicists (Feynman, Leighton, and Sands 1963b) Vol. 2, S 13-4, (Purcell and Morin 2013). The traditional formulation is strongly criticized with examples and analysis in (Purcell and Morin 2013), p. 500-507. (Purcell and Morin 2013) say: "This example teaches us that in the real atomic world the distinction between bound charge and free charge is more or less arbitrary, and so, therefore, is the concept of polarization density **P**." The concept of polarization density **P** density is central to the treatment of electrostatics and electrodynamics in Maxwell (Maxwell 1865b, 1865c; Whittaker 1951) and modern classics (Griffiths 2017), (Purcell and Morin 2013), (Zangwill 2013), and (Feynman, Leighton, and Sands 1963a), so the criticism of Purcell and Morin has great implications as it undercuts much of modern teaching. The parsing done here seems to resolve this problem while not deviating too much from the classical formulation, in my opinion.

The properties of charge and current in matter do not appear in the Ampere-Maxwell equation itself eq. (1), and equations derived from it, as we have seen. Rather the properties of charge and current need to be described by a separate conjoined theory (Eisenberg, Liu, and Wang 2022; Xu et al. 2023b, 2022) that shows how charge and current vary with conditions, particularly the electric field, just as the compressibility of matter must be included as a separate conjoined theory in fluid mechanics. Treating matter, charge, and electric/magnetic fields this way requires methods already well developed in fluid mechanics, particularly of complex fluids, that use the Energy Variational Approach developed by Chun Liu, more than anyone else.(Eisenberg, Liu, and Wang 2022; Xu et al. 2023b, 2022)

The classical form of the Ampere-Maxwell equation (1) involves the dielectric constant as one of the adjustable parameters of matter. The approximation of a dielectric constant as a single real positive number is a gross oversimplification, necessary at the time that Maxwell wrote his equations, but rather misleading in modern applications as we now document.

**Dielectric constant.** The assumption of a dielectric constant as a single real positive number is particularly inappropriate for ionic solutions that are not considered explicitly by (Purcell and Morin 2013), see (Eisenberg 2013; Barthel, Buchner, and Münsterer 1995; Barthel, Krienke, and Kunz 1998; Kremer and Schönhals 2003; Barsoukov and Macdonald 2018; Eisenberg 2019b)

Most of biology, and a great deal of chemistry occurs in ionic solutions, and so the approximation seriously limits the scope of use of the classical theories of electrodynamics. The range of concentrations of ions is large in electrochemistry and biology, with very large values found near the electrodes of batteries, proteins of ion channels and enzymes, and the nucleic acids of the genetic apparatus, values approaching the concentrations of 10 to even 100 molar found in solid crystals of salt (Eisenberg 2019b; Jimenez-Morales, Liang, and Eisenberg 2012; Eisenberg 2013). Crudely put, ion concentrations are often largest where they are most important in batteries, ion channels, nucleic acids, and many proteins. The dielectric properties of liquids depend on concentration. The approximation of a dielectric constant is a useful teaching tool and a necessary part of first investigations in ionic solutions and thus living systems but the approximation limits the scope of more realistic investigations (Eisenberg 2019b; Jimenez-Morales, Liang, and Eisenberg 2012; Eisenberg 2013).

The approximation of a single dielectric constant fails altogether when dealing with the remarkably diverse (and specific) dielectric properties of organic molecules revealed by their spectra, e.g., their infrared spectra. The spectroscopic properties of organics are so specific that they are used as fingerprints to identify individual chemical species.(Banwell and McCash 1994; Stuart 2005; Jaffé and Orchin 1962) The relation of dielectric and spectroscopic measures is explained by Parsegian (Parsegian 2006) p. 241-275. Dielectric and spectroscopic measures are different estimates of the same polarization currents. Dielectric and spectroscopic estimates were originally made at quite different frequencies, but that constraint came from technologies of previous centuries not from the molecules they measure.

The approximation of a single dielectric constant (a real number) was developed when measurements were confined to low frequencies, say < 100 Hz, as they were in Maxwell's time.

Modern data shows that polarization currents vary in a complex multi-dispersion way with frequency, type, and concentration of ions. All types of ions, not just the main charge carrier, change the dispersions.(Eisenberg 2013; Barthel, Buchner, and Münsterer 1995; Barthel, Krienke, and Kunz 1998; Kremer and Schönhals 2003; Barsoukov and Macdonald 2018; Eisenberg 2019b) Thus, a dispersion in a physiological salt solution will depend on many species, perhaps some in small concentrations and not on the main constituents. (The main constituents are typically $Na^+$ ions outside cells, and $K^+$ ions inside cells). The polarization current of an ideal dielectric, with a single dielectric constant—that is assumed in the classical Maxwell equations—cannot describe these properties.

It is interesting history that Maxwell anticipated the complexity of dielectric polarization quite explicitly (Maxwell 1865b): Vol. 1, p. 381. He was concerned that dielectric polarization might resemble electrochemical polarization, which modern experimentation confirms is the case.(Barthel, Buchner, and Münsterer 1995; Barthel, Krienke, and Kunz 1998; Kremer and Schönhals 2003; Barsoukov and Macdonald 2018)

The difficulties with the dielectric approximation are not confined to liquids. The approximation is barely adequate (Oriols and Ferry 2013; Eisenberg, Oriols, and Ferry 2022) even for crystalline solids like the silicon of our modern semiconductors. Currents in modern circuits operate at frequencies $> 10^8$ Hz in which bits of information might last $10^{-9}$ seconds. It is not reasonable to expect an approximation derived for frequencies $< 10^2$ Hz to work well at frequencies $> 10^8$ Hz. A single dielectric constant is often not good enough to allow design of new devices on the $10^{-9}$ second time scale, although it may be an adequate description of existing devices under some circumstances (Ayers 2018; Gielen and Sansen 2012; Gray et al. 2009; Hall and Heck 2011; Horowitz and Hill 2015; Scherz and Monk 2006; Sedra et al. 2020) and it is certainly an important teaching tool, and a necessary component of first investigations using 'toy' models. Later investigations require measurement and description of the properties of dielectrics and polarization currents under conditions of interest.

**Current Laws.** The Ampere-Maxwell equation (1) can be converted into a current law valid under the same conditions that the Ampere-Maxwell equation is valid. The Ampere-Maxwell equation is valid in essentially all circumstances, locations, and times that I know of.

Take the divergence of **curl B** and use the general mathematical result that divergence of the curl of a vector field is zero.

$$\text{Maxwell Current Law} \qquad \mathbf{div\ J}_{true} = \mathbf{div\ curl\ B} = 0 \qquad (2)$$

Equation (2) needs a name of its own because of its generality and so it is called the Maxwell Current Law in this paper. Equation (2) says in mathematics what Maxwell said in words. True current does not accumulate.

The fact that true current never accumulates allows understanding of important systems. In engineering, the lack of accumulation is the fundamental reason that Kirchhoff's law (and its generalizations) can be used on time scales faster than $10^{-10}$ seconds, even though it is usually derived and presented as a long-time approximation.

The fact that true current never accumulates allows understanding of important systems in biology. One example is the propagation of an action potential signal in nerve cells. Circuit laws have been used in traditional analysis of action potentials for a long time. (Hodgkin and Rushton 1946; Davis and de No 1947; Huxley 1972; Hodgkin and Huxley 1952) to quantify the electrical transmission of the signal.(Hodgkin 1937b, 1937a)

The transport of electrons, protons, and ions in mitochondria and active transporters (indeed, in proteins and enzymes in general (Eisenberg 1990)) are subject to the same physical laws as nerve cells and engineering circuits, even though the use of circuits is not traditional for these systems. The current laws are not confined to circuits as Maxwell himself made clear. (Maxwell 1865b), Vol.1, Sect.328, p. 377). Our analysis eq. (2) & (5) confirms his result. Current laws are true for three dimensional systems like mitochondria. Only vitalists believe that biological systems are immune from physical laws and constraints.

**True Current and Incompressible Fluids.** True current behaves like the flow of a perfectly incompressible fluid that Maxwell called the aether (sometimes called the absolute ether).

It is customary nowadays to avoid discussions of the aether itself (Kuhn 2012) although the current flow of the aether can be seen as light from the sun because of the displacement current it creates. The displacement current $\varepsilon_0\, \partial \mathbf{E}/\partial t$ allows electrodynamic waves to propagate through the near vacuum of space. It is customary today to deal with the current flow $\varepsilon_0\, \partial \mathbf{E}/\partial t$ of the aether, not the aether itself. The Michelson-Morley experiment convinced physicists that the movement of the aether could not be detected mechanically (Feynman, Leighton, and Sands 1963a), Vol 1, Section 15-3.

**Maxwell's True Current** is a property of the fields of electrodynamics and has useful properties that are independent of the properties of matter because the true current is does not accumulate with $\mathbf{div\ J}_{true} = \mathbf{0}$.

Many properties of a fluid flow that does not accumulate—that might be called 'incompressible' in loose language—are determined simply by the fact that it is incompressible (Zank and Matthaeus 1991; Christodoulou and Miao 2014). That may be why Maxwell called the total current a "true" current and emphasized that definition as his first equation in (Maxwell 1865a), eq. **A**, p. 465, 480.

It is not clear why Maxwell's focus on displacement current has been lost in modern literature, despite the vivid writings and presentations of Landauer (Landauer 1992). Maxwell called displacement current (and time dependence) one of the *"chief peculiarities [characteristics]"* of his work. Losing focus on displacement current leads to all sorts of difficulties when (for example) purely resistive circuits are considered. If circuits made of only resistors are perturbed by a step in potential or current at time zero, to a constant value, the analysis of textbooks (shown in numerous worked examples) gives results independent of time. Indeed, the eventual steady state responses observed in experiments are independent of time and are well described by the classical Kirchhoff analysis of eq. (6). Experiments, however, show prominent time dependent currents or voltages that are not described by classical Kirchhoff analysis, because it includes no time dependence at all. Specifically, if the resistor circuit is composed of roughly one megohm resistors, transients will be observed on the 0.1 to 1 microsecond time scale. These transients can be dealt with by modifying the circuits with stray capacitances, as we will discuss at length eq. (6)- (9), but the fact remains that treatments without displacement currents give results at odds with experiments. They have the basic time dependence implied by the Maxwell Ampere equation (1) itself.

The results of classical Kirchhoff analysis are valid only at steady state, but the examples and problems in the textbooks do not make that clear. Indeed, they use the classical Kirchhoff law to analyze circuits with time dependence when they turn to (for example) the high frequency circuits of our computer technology. The textbooks assume the student and reader will themselves know not to use the circuits (as drawn in idealized form) but will know to add in the supplementary circuit elements needed to introduce time dependencies. In my experience, this assumption puts a difficult burden on students and leads to confusion between the role of idealized theory, realistic simulations, and actual experimental results.

**Conduction current is compressible.** Other kinds of current—like the conduction current **J**—accumulate. Those currents **J** depend on the properties of charge density as it accumulates and thus on the properties of matter (through the continuity eq. (4)) as well as the properties of electrodynamic fields.

Conduction current requires consideration of macroscopic numbers of charges and their interactions because the continuity equation requires such consideration. True current does not require consideration of charges, for most purposes. Consider the design of the circuits in general including the circuits of our computers (Bush and Wiener 1929; Tuttle 1958; Ghausi and Kelly 1968; Guillemin 1931, 2013; Balabanian and Bickart 1969; Weinberg 1975; Valentinuzzi and Kohen 2013; Bhat and Osting 2011; Eisenberg 2022b, 2023; Ayers 2018; Gielen and Sansen 2012; Gray et al. 2009; Hall and Heck 2011; Horowitz and Hill 2015; Scherz and Monk 2006; Sedra et al. 2020). Consider the action potential of nerve and muscle. These do not require analysis of charges for most purposes.(Huxley 1972)

**The physics of the Maxwell Current Law** should not be hidden by the mathematics of the **div** and **curl** operators of vector calculus. The Maxwell Current Law (2) arises because the electric and magnetic fields created by electrodynamics move charges and atoms. Combined with the displacement current $\varepsilon_0\, \partial \mathbf{E}/\partial t$, the charge movements guarantee that the total current does not accumulate.

The total current does not accumulate with surprising consequences (Eisenberg 2020a) particularly for single file transport in one-dimensional systems as stated clearly by (Landauer 1992) focusing on physical systems. The conduction current in one dimensional biological systems has complex properties.(Hodgkin and Keynes 1955; Hille and Schwartz 1978; Hille 2001) The total current in one dimensional systems does not. If total current flows in one dimension, it is equal everywhere and at every time. That is the mathematical consequence of the Maxwell-Ampere equation. Biological properties that depend only on total current are much easier to analyze than those that depend on the movement of particular ions for that reason. The extensive analysis of hopping—that forms much of the classical literature of electron transport (Landauer 1992) and channel biology (Hodgkin and Keynes 1955; Hille and Schwartz 1978; Hille 2001)—is not needed if the output variable of interest is total current. That analysis in biology was based on a mechanical model of the interaction of spheres ('billiard balls'), without charge, ignoring electrical forces, because mechanical analysis was thought to be much simpler than analysis of charge motion (personal communication AL Hodgkin and separately RD Keynes to Bob Eisenberg) although ions are highly charged. It turns out, however, that analysis of the current of the charges is much simpler than the analysis of uncharged billiard balls, if the current analyzed is the total true current. If total current flow is confined to one dimension in a channel, it is equal everywhere at any particular time. The flows of ions are much more complicated, indeed, as a glimpse of the large modern literature shows.(Eisenberg 2020a; Kopec et al. 2018; Luchinsky et al. 2014; Köpfer et al. 2014; Kraszewski et al. 2007)

**Conduction Current Accumulates.** The conduction current **J** does accumulate even though $\mathbf{J}_{true}$ does not.

$$\mathbf{div\,J} = -\mathbf{div}\,(\varepsilon_0\,\partial\mathbf{E}/\partial t) \qquad (3)$$

The divergence of **J** is zero when $\varepsilon_0\,\partial\mathbf{E}/\partial t = 0$.

Indeed, the continuity equation (4) shows that conduction current **J** accumulates as charge density $\rho$ changes.

$$\mathbf{div\,J} = -\,\partial\rho/\partial t \qquad (4)$$

The continuity equation (4) can be derived from the Maxwell-Ampere Equation (1), and a form of Gauss' law (Feynman, Leighton, and Sands 1963a); eq. **G** of (Maxwell 1865a), p. 465 and 485). We reiterate for clarity that in our nomenclature **J** includes the polarization current of an idealized dielectric $(\varepsilon_r - 1)\varepsilon_0\,\partial\mathbf{E}/\partial t$. **J** excludes only universal displacement current $\varepsilon_0\,\partial\mathbf{E}/\partial t$. **J** is the conduction current of any charge with mass, however brief, small, or transient, including polarization current of matter.

**Conduction Current Accumulates as Charge Density.** Integrated circuits contain too many charges to analyze with the continuity equation (4). Rather, integrated circuits are analyzed by the Kirchhoff current law eq. (5), without explicit consideration of charges.(Ayers 2018; Gielen and Sansen 2012; Gray et al. 2009; Hall and Heck 2011; Horowitz and Hill 2015; Scherz and Monk 2006; Sedra et al. 2020) It is easier to deal with currents than charges. The number of currents that must be considered is much less than the number of charges in macroscopic systems that often contain an incomputable number of charges and charge interactions.

Conduction current is also harder to analyze than total current because conduction current depends on the accumulation of charges that diffuse and flow by convection. Those processes involve many parameters including the details of the structures and boundaries in which diffusion and convection occur. Such complexity is particularly hard to deal with in biological systems like nerve fibers where many types of ionic currents flow in different channels, and components of cells and the currents are driven by diffusion and migration, and sometimes convection. The complexity is hard to deal with in mitochondria where electrons, protons, and ions flow, each carrying significant current important in the generation of ATP, the main function of mitochondria. Despite this complexity it is in fact possible to use conservation laws and descriptions of structure and structural components to compute currents directly comparable to experiments. (Eisenberg 2022c; Xu et al. 2023a, 2023b; Song et al. 2022; Zhu et al. 2021b, 2021a; Zhu et al. 2019; Xu et al. 2018)

**Engineers rarely analyze charge.** Many scientists—including me for most of my life—have thought that knowledge of the locations and properties (and accumulation) of charges is needed to deal quantitatively with circuits or electromagnetic phenomena including circuits. Feynman says nearly as much, for example, in (Feynman, Leighton, and Sands 1963a), Vol 2, Section 13-4. However, engineers have built and used circuits without explicitly studying charges for a long time, for some 160 years since the introduction of the telegraph, using just the Kirchhoff current law eq. (5) that only involves currents and does not deal directly with charges. They have shown how to design successful circuits that bring power and information to our homes and businesses without detailed knowledge of the distribution of charges.(Bush and Wiener 1929; Tuttle 1958; Ghausi and Kelly 1968; Guillemin 1931, 2013; Balabanian and Bickart 1969; Weinberg 1975; Valentinuzzi and Kohen 2013; Bhat and Osting 2011; Eisenberg 2022b, 2023; Ayers 2018; Gielen and Sansen 2012; Gray et al. 2009; Hall and Heck 2011; Horowitz and Hill 2015; Scherz and Monk 2006; Sedra et al. 2020)

**Circuits are the main use of electricity in our world.** Circuits carry power to our homes and businesses, lighting the night; they carry information everywhere, whether as sounds in telephones, images in videos, or bits and bytes in our computers.

Circuits are designed by laws that describe the flow of currents, chiefly Kirchhoff's current law (5). Circuit laws like eq. (2) and (5) are needed to define circuits precisely. Circuits can be defined without explicit discussion of charges and have been since the invention of the telegraph.

Biologists do not always recognize the importance of circuits perhaps because circuits hardly appear in traditional biochemically oriented textbooks.(Alberts et al. 1998; Berg, Tymoczko, and Stryer 2010) Circuits appear in the traditional literature only when dealing with nerve and muscle cells.(Hodgkin and Rushton 1946; Davis and de No 1947; Huxley 1972; Hodgkin and Huxley 1952) The cable equation was widely known to workers in Cambridge UK in the 1930's in the afterglow of the triumphant technology of Victorian England including the trans-Atlantic cable. 1850 (Kelvin 1855, 1856) (Gordon 2008).That history and physics/engineering is not widely known to biologists today and the importance of the cable equation has been more or less lost to the common knowledge of molecular and cellular biologists.

**Current is not confined to circuits.** It is important for biology that the Maxwell Current Law eq. (2) does not require current to be confined to circuits. The Maxwell Current Law is true whenever classical electrodynamics are true. Maxwell ((Maxwell 1865b), Vol.1, Sect.328, p. 377) analyzes systems at some length that are not circuits. Maxwell solves electric field problems that are not circuits in explicit detail including boundary properties and conditions, showing how the time dependent solutions of those boundary value problems depend on the displacement current $\varepsilon_r \varepsilon_0\, \partial \mathbf{E}/\partial t$. If the displacement current is not included, the solutions to the problems are incorrect, as Maxwell makes quite clear. Indeed, he uses this fact to justify his insistence on the importance of using true current to estimate the 'total movement of electricity'.

**Kirchhoff Current Law** is used throughout engineering and physics to define the properties of circuits. The current used is usually defined as the flux of charges and does not include displacement current or a time derivative. Examples and problems in the textbooks show this fact without ambiguity. Examples containing only resistors are common and they do not contain time derivatives even if the potentials and currents involved vary with time.

The examples of textbooks unfortunately differ from reality, it must be realized. When circuits are actually constructed from these resistors, and potentials or currents are applied at time zero, time dependence does appear in responses even though it is not found in the examples of the textbooks. The textbooks only analyze the steady state response which of course does not depend on time. That is natural because the Kirchhoff law of currents that they use is a low frequency approximation. But an experiment using steps in applied potential or current shows transients. Specifically, resistor networks made of roughly one megohm resistors contain many transients on time scales of 0.1 to 1 microsecond. Transients appear in experiments as Maxwell's discussion shows they must: Vol. 2, Section 610, p. 232 (Maxwell 1865b). Engineers are aware of this problem and use fixups to correct it. The dielectric (polarization) current—needed to make the time independent Kirchhoff Law fit time dependent experimental data—is usually provided by fictitious supplements to the circuit in question called 'stray capacitances'. The dielectric (polarization) current is needed to make the (time independent) Kirchhoff Law (5) fit experimental data that exhibits time dependence.

Kirchhoff Current Law eq. (5):

**Kirchhoff Current Law for Fields** $\quad\mathbf{div\,J\,=\,0}\quad$ **J** is conduction current $\quad$ (5)

Kirchhoff Current Law eq. (5) is a special case of the Maxwell Current Law eq.(2). This form of Kirchhoff's law is unfamiliar because it does need not be inside a circuit. **J** is not the flux of charge. **J** is the conduction current defined above that includes polarization currents and currents of ideal dielectrics.

Kirchhoff's law in texts of circuit design and analysis is often said more clearly in words than in equations: 'all the currents that flow into a node—positive quantities—flow out as negative quantities". (Bush and Wiener 1929; Tuttle 1958; Ghausi and Kelly 1968; Guillemin 1931, 2013; Balabanian and Bickart 1969; Weinberg 1975; Valentinuzzi and Kohen 2013; Bhat and Osting 2011; Eisenberg 2022b, 2023; Ayers 2018; Gielen and Sansen 2012; Gray et al. 2009; Hall and Heck 2011; Horowitz and Hill 2015; Scherz and Monk 2006; Sedra et al. 2020) Or 'the sum of all currents at a node is zero'. The meaning of 'Currents' is usually not discussed in the texts of circuit design. Kirchhoff's law is usually considered a low frequency approximation in those texts and time derivatives and displacement currents are not discussed in the application of Kirchhoff's law to circuits containing only resistors, for example.

**<u>Kirchhoff's Law, Classical and True Current</u>**. As far as I know, derivations of Kirchhoff's law in physics and electronics texts do not include the displacement current term explicitly as part of the current used in the law. Examples involving only resistors illustrate this point without ambiguity. They do not include derivative terms or displacement current if the circuit analyzed does not include capacitors or inductors. Displacement current is handled by fictitious supplementary components $\varepsilon_r\,\varepsilon_0\partial\mathbf{E}/\partial t$ called 'stray' or 'parasitic' capacitances as mentioned previously, discussed at length later in this paper, see eq. (7)-(9).

Classical treatments of Kirchhoff's current law that do not use Maxwell's true current appear to violate Maxwell's statement: true current "*must be considered in estimating the total movement of electricity*" *(loc.cit.).* When the true current is used in a current law, fictitious supplementary components are not needed to satisfy the Maxwell-Ampere equation.(Eisenberg et al. 2018) To say it crudely: "Kirchhoff's law can be exact" (Eisenberg 2019c) if true current is used.

Supplementary components can be confusing, but they are not wrong because they provide both the mathematically necessary (and physical required) universal displacement current $\varepsilon_0 \partial \mathbf{E}/\partial t$ (Eisenberg 2022b) and supplementary components $(\varepsilon_r - 1)\varepsilon_0 \partial \mathbf{E}/\partial t$ often needed to describe material properties. (Eisenberg, Oriols, and Ferry 2017; Eisenberg, Oriols, and Ferry 2022)

Kirchhoff's law is customarily applied to ideal circuits that do not explicitly contain stray capacitances. For each node $i$ in the circuit, one sums the conduction currents from the $k$ branches attached to that node. $J_{i(k)}$ is usually defined as the flow of electrons. In most treatments, displacement current $\varepsilon_r \varepsilon_0 \partial \mathbf{E}/\partial t$ is not discussed as a possible component of $J_{i(k)}$. Time derivatives are not found in textbook examples made of only resistors. The lack of time derivatives illustrates in undeniable detail that the textbook authors do not mean to include displacement current in the current of Kirchhoff's law.

| **Classical Kirchhoff Current Law for $i^{th}$ Node in Circuits** | $\sum J_{i(k)} = 0$ | $J_i$ conduction current | (6) |

The classical Kirchhoff Current Law (5) as usually presented appears to be inconsistent with Maxwell's statement "*that the time-variation of the electric displacement, must be considered*". (Maxwell 1865b) Vol. 2, Section 610, p. 232. Eq. (6) can be modified to contain the displacement current, as we shall see, eq.(9), thereby becoming consistent with eq. (2). The Kirchhoff law eq. (5) and (6) can be made physically consistent with the Maxwell Current Law eq. (2), and thus the Ampere-Maxwell eq. (1) by supplementing circuits with fictitious stray capacitors, as we show with eq. (7)-(9) and further discussion.

**Paradigm Change.** We consider in more detail the ways to reconcile the Maxwell and Kirchhoff Current laws, eq. (2) and (5), because of the general importance of the subject, and the novelty of my approach. Reconciliation is needed because the classical Kirchhoff law does not include the displacement current.

A simple logical manipulation of the Maxwell quotation in the beginning of this paper reads "It is incorrect to compute the movement of electricity without the displacement current." That is to say, according to Maxwell, it is incorrect to compute the movement of electricity without the displacement current. This statement raises a problem because the movement of electricity is commonly analyzed without displacement currents in the classical applications of Kirchhoff's current as we have just discussed. It is not clear why displacement currents are not included in textbook treatments. The reasons seem social and psychological more than mathematical or scientific.(Kuhn 2012)

One way to reconcile the Maxwell and Kirchoff Current laws is by approximation, viewing the classical Kirchhoff law as a low frequency law described in detail below, eq. (7)-(9)

The other way to reconcile is to redefine current, using Maxwell's definition of true current, eq. (1) & (2), an approach I prefer (Eisenberg 2019c; Eisenberg et al. 2018)

(1) because it automatically shows why the current law works at high frequencies,
(2) because it automatically applies to three dimensional systems that are not obviously circuits and
(3) because it arises naturally from the fundamental equations of electrodynamics.
(4) because it emphasizes the reality that large (displacement) currents flow in a vacuum devoid of charge (or conduction current) as the currents $\varepsilon_0\, \partial \mathbf{E}/\partial t$ create magnetic fields and propagate electromagnetic radiation like light.

**Reconciliation by Approximation.** When conduction current is much larger than displacement $\varepsilon_0\, \partial \mathbf{E}/\partial t$, true current is mostly conduction current. In that case, at low frequencies, the usual Kirchhoff Current Law (6) may be good enough, helpful even when dealing with microwave circuits (Okoshi 2012; Schwierz and Liou 2003; Fukunaga and Kurahashi 2007; Mei 2000), if used with caution and care. The Kirchhoff Current Law eq. (6) is viewed as an approximation to the Maxwell Current Law (2) in most textbooks and derivations for that reason, in my view.(Bhat and Osting 2011)

Care is important, however, even when the displacement term is small, because the initial condition associated with the time derivative $\partial \mathbf{E}/\partial t$ is lost in the approximation that $\partial \mathbf{E}/\partial t = 0$. Losing initial conditions causes difficulties in almost any system described by differential equations, as studied at length in singular perturbation theory (Kevorkian and Cole 1996).

Losing the initial condition in electrical systems can lead to paradoxes that make the role of charge hard to understand. If $\varepsilon_0\, \partial \mathbf{E}/\partial t$ is set to zero, all the conduction current that flows into a resistor also flows out of the resistor, so one must wonder how the charge that drives the conduction current accumulates in the first place. The charge arises, of course, in an initial condition that is not included in a treatment like eq. (5) where $\partial \mathbf{E}/\partial t = 0$. The initial charge can be included as extra information added to supplement Kirchhoff's law, see eq.(7). These fix-ups are subtle and confusing (even in outstanding textbooks (Desoer and Kuh 1969)) but they are not wrong as long as somehow or other they introduce the universal displacement current $\varepsilon_0\, \partial \mathbf{E}/\partial t$ and concomitant time dependence. Even when correct, however, the fixups often mystify students and scientists from other fields. The classical use of a long-time approximation to design circuits that switch in nanoseconds seems artificial and confusing (and of course not unique) to most students and scientists, in my experience.

**Reconciliation with Stray Capacitance.** As we have mentioned, the classical way to reconcile the Maxwell and Kirchhoff current laws eq. (2) & (6) is to change the system that the laws describe. Virtual components are almost always added to idealized circuits of textbooks (Horowitz and Hill 2015). Idealized circuits are almost always revised as they are converted to actual circuits on circuit boards. 'Stray capacitances' are almost always added that do not actually exist as isolated elements—i.e., capacitors—in the real circuit (Bush and Wiener 1929; Tuttle 1958; Ghausi and Kelly 1968; Guillemin 1931, 2013; Balabanian and Bickart 1969; Weinberg 1975; Valentinuzzi and Kohen 2013; Bhat and Osting 2011; Eisenberg 2022b, 2023; Ayers 2018; Gielen and Sansen 2012; Gray et al. 2009; Hall and Heck 2011; Horowitz and Hill 2015; Scherz and Monk 2006; Sedra et al.

2020; Lienig and Scheible 2020). These extra components reconcile Maxwell and Kirchhoff current laws (2) & (6).

The stray capacitances added are often of the form $\varepsilon_r\,\varepsilon_0 \partial \mathbf{E}/\partial t$. These added capacitances allow circuits to describe coupling to nearby structures in the circuit layout, as well as the usually smaller but unavoidable displacement current $\varepsilon_0\,\partial \mathbf{E}/\partial t$, see eq.(9). The extra components $\varepsilon_r\,\varepsilon_0 \partial \mathbf{E}/\partial t$ allow circuits to deal with reality quite well at high frequencies, reaching to microwave frequencies in favorable cases.(Okoshi 2012; Schwierz and Liou 2003; Fukunaga and Kurahashi 2007; Mei 2000) They appear as extra terms (see eq. (9)) in addition to those included in textbook low frequency formulations of the Kirchhoff Current Law eq. (6). Successful circuit designs usually place 'stray capacitances' $\varepsilon_r\,\varepsilon_0 \partial \mathbf{E}/\partial t$ between all nodes in a circuit. The examples and problems in textbooks need to be modified to include these stray capacitances if they are to describe the actual properties of real circuits. Students and scientists from other fields who do not know how to add stray capacitances are discouraged, as well as mystified, by the difference between actual results and the behavior of their idealized circuits in my experience.

**Far distant ground.** Successful circuit designs often place 'stray capacitances' between the nodes and a far distant zero potential ground as well as between nodes. The far distant ground is the typical far field boundary condition in circuit theory. It is the typical boundary condition assumed in many electrostatic problems including those of classical statistical mechanics, citations in (Eisenberg 2020c; Eisenberg 2022a)

A *single* far field boundary condition 'at infinity' is incompatible, however, with a general treatment of electrodynamics as explained at length in (Eisenberg 2022a): the wave properties of the Maxwell equations and also the complex structures of engineering and biological systems prohibit such simplicity. One must use an explicit limiting process, specific to particular applications, to define far field boundary conditions even if the conditions are far away, "at infinity". Otherwise, results are not unique and cannot be computed with results reproducible from one research group to another.

**Stray capacitances are not usually included** in idealized circuits because the capacitances depend on the layout of the circuit—the way it is wired up—and the layout is not shown in idealized circuit diagrams.

The conversion of idealized circuits into functional designs is a discipline of its own, of great importance in modern design of integrated circuits (Eisenberg, Oriols, and Ferry 2017; Eisenberg, Oriols, and Ferry 2022) because stray capacitances and other non-ideal effects increase in importance as the density of devices increases, and the size of devices approach atomic dimensions and the time scale of circuit operation decreases.(Hall and Heck 2011; Ayers 2018; Lienig and Scheible 2020; Howe and Sodini 1997; Scherz and Monk 2006; Gielen and Sansen 2012) The values of stray capacitances are determined empirically in that design process.(Hastings 2001; Lienig and Scheible 2020)

**Reconciliation by Redefinition.** As we have seen, another way to reconcile the Kirchhoff Current Law eq. (6) and the Maxwell Current Law (2) is simply to use a true current in the law, as Maxwell instructs in the quotation that begins this paper and we advocated before we knew of his instruction.(Eisenberg 2019c; Eisenberg et al. 2018)

The exact reconciliation by redefintion simply changes the definition of current (in the traditional Kirchhoff's law for circuits) to true current $\mathbf{J}_{true}$. That redefined current is Maxwell's true current eq. (1), of the Ampere-Maxwell differential equation for the magnetic field. The true current $\mathbf{J}_{true} = \mathbf{J} + \varepsilon_0\, \partial \mathbf{E}/\partial t$ must include the displacement current $\varepsilon_0\, \partial \mathbf{E}/\partial t$ because true current must flow in a vacuum if light is to propagate as an electrodynamic wave in the near vacuum of the space between the sun and earth. $\mathbf{J} = \mathbf{0}$ in a vacuum.

**<u>Three dimensional structures are different</u>.** For three dimensional structures, the story is different. For mitochondria, nerve, and cardiac cells, and other three-dimensional structures, it is not always clear how to approximate the three-dimensional layout of biological systems with a circuit model. Indeed, the circuit representation of such complex structures is not unique except in the cases where the cable equation can be derived from the general equations of electrodynamics (as discussed later) or verified experimentally. (Hodgkin and Rushton 1946; Barcilon, Cole, and Eisenberg 1971),(Mobley, Leung, and Eisenberg 1974, 1975; Kevorkian and Cole 1996) pp 218-238.

Circuit models are in fact not used in the classical literature of mitochondria or the chemiosmotic theory, although figures showing arrows and the movements of electrons and pathways for those movements are commonplace. The arrows are not identified as currents. The flows of electrons exist in this classical work, as in real mitochondria, but the laws of current flow, or electrodynamics have not been used to describe those currents, as far as I know, except in the papers of our research group (Xu et al. 2023b, 2023a).

If currents are viewed as fluxes of charge, the electron flows in classical chemiosmotic models are indeed tricky to interpret and the fluxes of protons are hardly easier. But if the currents are viewed through the lens of electrodynamics, interpretations are easier, as physicists have found for a long time, since the invention of the telegraph, and the introduction of the Maxwell equations. Viewed as true currents, including displacement current, the arrows in classical models of mitochondria and the chemiosmotic hypothesis have a specific meaning and follow quantitative laws well established in physics. Clarity emerges when the equations of physicists replace the words of biologists.

Using Maxwell's definition of current makes the current laws general. The Maxwell Current Law (2) does not require discussion of circuits, as Maxwell himself demonstrated. (Maxwell 1865b), Vol.1, Sect.328, p. 377.

The true current can be used to describe the following.
  (1) electron flows in circuits,
  (2) current flows in mitochondria, including
  (3) proton flows and
  (4) trans-membrane electron flows and
  (5) horizontal electron flows, and
  (6) ionic flows in nerve including
  (7) ion flows in membranes or
  (8) ion flows in cytoplasm,

The flows are carried by whatever ions are present (Eisenberg et al. 2018; Eisenberg 2019c) as can be shown experimentally (Eisenberg et al. 2018; Eisenberg 2019c; Chandler, Hodgkin, and Meves 1965; Baker, Hodgkin, and Shaw 1962b, 1962a) even in cells as complicated as skeletal muscle.(Mobley, Leung, and Eisenberg 1974, 1975)

Maxwell's definition of current is easy to use and does not involve approximation. It requires a slight change in the treatment of existing idealized circuits or analyses of field problems in applied mathematics that use the dielectric constant approximation, as they usually do.(Eisenberg 2022b) The use of Maxwell's definition of current represents a paradigm shift and so is difficult for scientists to implement beyond the logical issues involved.(Kuhn 2012)

**Laplace Transforms.** Analysis of circuits usually uses Laplace transforms because most circuits of technological interest are linear systems. Nonlinear systems are often analyzed beginning with a linearization. Nonlinear technological devices are usually analyzed by linearization around the operating points where they function. The operating point is rarely a point of zero flux or equilibrium in engineering systems, although it is often a zero point in the linearizations of statistical mechanics.

Maxwell's definition of current in eq. (1)&(2) is easily formulated using Laplace transforms. The Laplace transform $\widehat{\mathbf{J}}_{true}(s)$ of the true current $\mathbf{J}_{true}(t)$ of eq. (1) is

**Laplace transform of true current** $\qquad \widehat{\mathbf{J}}_{true}(s) = \widehat{\mathbf{J}}(s) + \varepsilon_0 s \widehat{\mathbf{E}}(s) - \varepsilon_0 \mathbf{E}(t=0)$ (7)

Here the symbol $s$ is the Laplace transform frequency variable, Laplace transforms are indicated by a 'hat' as in the Laplace transform of the electric field $\widehat{\mathbf{E}}(s)$. If we set initial conditions $\mathbf{E}(t)$ to zero $\mathbf{E}(t=0)$, as is customary in circuit analysis and synthesis, every modified current in a circuit contains an extra term like $\mathbb{C}s\widehat{\mathbf{E}}(s)$, where the universal effective capacitance $\mathbb{C} = \varepsilon_0$ is simply the electrical constant, independent of the properties of matter. $\mathbb{C}s\widehat{\mathbf{E}}(s)$ is Maxwell's displacement current, the term that is essential for the understanding how electromagnetism in a vacuum supports a propagating wave like light.

Circuits analyzed with the Maxwell Current Law (2) will thus contain many effective capacitances $\mathbb{C}s\widehat{\mathbf{E}}(s)$ that include the universal displacement current $\varepsilon_0\, \partial \mathbf{E}/\partial t$, one for each circuit element and wire. Combined, the effective capacitances will form an inescapable part of the stray capacitances.

Saying the same thing in equations, we rewrite eq. (2) for the $i^{th}$ node—with $k$ branches attached—in a circuit using the Laplace transform variables of eq. (7) with zero initial conditions everywhere.

$$\sum \widehat{\mathbf{J}}_{true;i(k)}(s) = \sum \left( \widehat{\mathbf{J}}_{i(k)}(s) + \varepsilon_0 s \widehat{\mathbf{E}}_{i(k)}(s) \right) = 0 \qquad (8)$$

This equation (8) can be rewritten so we recognize the *stray* capacitance currents $\varepsilon_0 s \widehat{\mathbf{E}}_{i(k)}(s)$ added to the branch currents $\widehat{\mathbf{J}}_{i(k)}(s)$ of the classical Kirchhoff Law eq. (6).

$$\underbrace{\sum \hat{J}_{true;i(k)}(s)}_{\text{Maxwell Law}} = \underbrace{\sum \hat{J}_{i(k)}(s)}_{\text{Kirchhoff Law}} + \underbrace{\sum (\varepsilon_0 + \textbf{\textit{Layout}}) \cdot s\hat{\textbf{\textit{E}}}_{i(k)}(s)}_{\text{Stray Capacitances}} = 0 \qquad (9)$$

The last few equations show how Kirchhoff's law eq. (6) can create an exact description (Eisenberg 2019c) of what flows through a resistor (Eisenberg et al. 2018).

Eq. (9) includes the 'Layout' capacitance term $(\textbf{\textit{Layout}}) \cdot s\hat{\textbf{\textit{E}}}_{i(k)}(s)$ that arises from geometric coupling to nearby structures. These material stray capacitances are often much larger than the unavoidable universal vacuum term $\varepsilon_0 \cdot s\hat{\textbf{\textit{E}}}_{i(k)}(s)$.

**Current in Biological Systems.** Kirchhoff's Current Law as usually formulated eq.(6) cannot be used with confidence in biological systems like mitochondria because their circuit representation has not been established and is unlikely to be unique. Mitochondria are not circuits.

The Kirchhoff circuit formulation can be used successfully in nerve fibers because they have been shown experimentally to behave like circuits (Hodgkin and Rushton 1946), and the circuit representation has actually been derived from one of the field equations of electrodynamics, the Poisson equation.(Barcilon, Cole, and Eisenberg 1971), (Kevorkian and Cole 1996) pp 218-238.

Hodgkin, Huxley and Katz used the properties of displacement current to validate the results of their voltage clamp experiments see Fig. 10 and eq. 11 of (Hodgkin, Huxley, and Katz 1952). That analysis provided important motivation (personal communication, AL Hodgkin to Bob Eisenberg, ~1970) for them to continue their initial voltage clamp experiments into a second season at Plymouth (Huxley 1972) and motivation (personal communication AF Huxley to Bob Eisenberg,~1970) for Huxley to perform numerical calculations of the action potential with a hand driven calculator that could only add and subtract (Hodgkin and Huxley 1952). Both Hodgkin and Huxley were fearful that membrane processes or conductances that inherently depended on $\partial \textbf{E}/\partial t$ could not be analyzed by the voltage clamp, in which $\partial \textbf{E}/\partial t = 0$ or infinity but nothing else.

The total current across all the membranes of a short system like a mitochondrion must sum to zero.(Barcilon, Cole, and Eisenberg 1971) The total current includes the membrane capacity current—a displacement current $\varepsilon_m \varepsilon_0\ \partial \textbf{E}/\partial t$—and the conduction currents carried by ions, electrons, and protons. If the Maxwell redefinition of current is used, the displacement current is automatically present. If the classical definition of current is used, care must be taken to include the displacement current. This is not just a theoretical issue.

A short organelle like a mitochondrion has a single internal potential in the matrix within the mitochondrion, almost independent of spatial location (Barcilon, Cole, and Eisenberg 1971), and the membrane currents, arising in the lipid membrane, and the channels, and protein complexes in the mitochondrial membrane, has nowhere else to go. There is no long axon to conduct the current away, as in a nerve fiber (Kevorkian and Cole 1996) pp 218-238. The Maxwell and Kirchhoff Current Laws (2) & (6) combine with the shape and size of the mitochondria to force the total current of all types across all membranes to sum to zero.

**Physical Basis of the Maxwell Current Law.** The total currents across the mitochondrial membrane, and the total currents linking proteins complexes, are the true currents of Maxwell and follow the Maxwell Current Law (2), entirely independent of the nature of the charges that carry the currents, whether electrons or protons or ions or the material displacement current of the lipid membrane or the universal displacement current of the vacuum, for that matter.

The physical basis of the Maxwell Current Law (2) lies in the electrical and magnetic fields produced by the Ampere-Maxwell equation (1) and the other Maxwell partial differential equations. ***It should clearly be understood that the presence of a time derivative in the Ampere-Maxwell equation ensures that a time derivative must be included in current laws themselves if they are to be consistent.*** Removing the time derivative term is not justified mathematically. It is justified physically only when shown to be appropriate and accurate enough.

One of the Maxwell equations says the divergence of the magnetic field is zero. The divergence is zero because no magnetic monopoles exist (comparable to the electrical monopoles the electron or proton). The Ampere-Maxwell equation (1) is the only source of the magnetic field, when the boundary conditions of the equation are included in the definition of the equation (1), as they must be if the equation is to have solutions.

The Ampere-Maxwell equation (1) guarantees that the electric and magnetic fields take on values that prevent the accumulation of true current. The total current across the mitochondrial membrane does not accumulate. The total current flowing from one protein complex to another, linking them into a functional unit, does not accumulate, even if the conduction components of that linking currents do accumulate. That is the physical counterpart to the mathematical derivation of the Maxwell Current Law given previously.

It is important to reiterate that the classical current is different from total current in important ways. That is probably why Maxwell called total current 'true' current. Classical current does accumulate because it is only the conduction current. ***Maxwell clearly indicates that only true current allows understanding of the movement of electricity.*** Classical current can be used but only if it is supplemented by the displacement current and the continuity equation. A derivative term must be included one way or another, to ensure consistency with the Maxwell equations, and more importantly, agreement with experimental data.

The classical conduction current accumulates according to the properties of charge accumulation (i.e., the continuity equation (4)). Those properties can only be calculated:
(1) by studying all the ions present, which is if the order of $10^{15}$ and
(2) by studying the interactions of the ions. Pairwise interactions number some $10^{15}$ factorial! Coarse graining is needed when dealing with numbers like this that are uncomputable, nearly in principle. It is fortunate that coarse graining can be provided by the Maxwell-Ampere law (1) and is thus exact.

**Coarse graining can be exact.** Coarse graining in simulations and applied mathematics usually involves approximation and is not unique. Indeed, it often is a matter of dispute between scientists using different approximations that are hard to compare, let alone reconcile. In the case we consider here, however, the coarse graining can be exact if the coarse graining is the Maxwell Ampere law equation (1). That coarse graining can be exact because the law actually invokes additional physics that make it exact—the properties of displacement current arising from the Maxwell-Ampere equation. Those properties exist everywhere and guarantee that total current does not accumulate anywhere, making the number of true currents that need to be evaluated in a system very much less indeed than the number of charges. The number is typically hundreds although of course it is very variable. The number of calculations of charges is much larger than $10^{15}$ because macroscopic flows, driving forces *and interactions* are involved.

**Maxwell Coupling.** The fluxes of ions and electrons across membranes are coupled by their summation in systems like short organelles, like mitochondria. The fluxes of ions and electrons across membranes cannot be independent in such systems. The fluxes in all five respiratory complexes—as well as all other membrane proteins and lipids—are coupled because they have a definite sum, zero, under all conditions. If one membrane conduction current increases, the other must decrease, or be balanced by a membrane capacity current $\varepsilon_m \varepsilon_0 \, \partial \mathbf{E}/\partial t$, to keep the sum zero, as required by eq. (2). If current flows in one respiratory complex, it changes current flow in other complexes. From a circuit viewpoint, current in one component of a circuit changes current in another component. The spatial distribution of electrical potential changes to force this change in current, just as the spatial distribution in potential changes in a circuit changes as currents change. From a physics point of view, total (true) current does not accumulate.

From a classical biochemical point of view, with its focus on charge, these couplings are not easily seen. They are not mentioned in the classical chemiosmotic theory and simulations as far as I know. It is difficult to add up $10^{15}$ charges accurately enough to see the couplings that "stare you in the face" when considering currents. Those couplings are so powerful that they have provided the circuits of our technology, from the telegraphs of 1840, before the atomic basis of charges was considered, long before the movement of atomic charges could be computed in simulations of molecular dynamics. The couplings are powerful enough to couple currents on opposite sides of the Atlantic Ocean, from Ireland to Newfoundland, in Kelvin's Atlantic cable. (Hodgkin and Huxley 1952; Hodgkin and Rushton 1946; Davis and de No 1947; Kelvin 1855, 1856) The couplings are universal enough to allow design of the circuits of our computers with bits of information some $10^9 \times$ briefer than the Morse code of telegraphs. The couplings allow the understanding of the nerve action potential. Perhaps the couplings are used by mitochondria in the production of ATP.

The Maxwell coupling produced by summation is enforced by electrodynamics whether the currents are carried by ions, or electrons, or displacement, no matter how the currents vary with time or the driving forces that create them as we have discussed. The Maxwell Current Law (2) applies to currents driven by chemical reactions, protonmotive forces, electrochemical potentials, convection, heat flow, and so on. The coupling can be embedded in models with the structural complexity of biological systems, albeit oversimplified. (Eisenberg 2022c; Xu et al. 2023a, 2023b; Song et al. 2022; Zhu et al. 2021b, 2021a; Zhu et al. 2019; Xu et al. 2018)

**Biological systems are not immune from physical laws.** The laws of electrodynamics apply to biological systems, including the electron transport chain of mitochondria, and the mechanism of action potential propagation. In mitochondria, the physical laws enforce the discipline of coupling of conduction currents in all the respiratory complexes (and membrane proteins and lipids) of the mitochondrion. The discipline is enforced by the changes in electric field that occur and the conduction currents that change as the field changes.

**Local Coupling.** The membrane and linking currents in mitochondria are unavoidably coupled by physics as we have just discussed. The conduction currents are coupled locally as well, by chemistry. Chemical reactions within the protein complexes in and near the membrane produce coupling. Flows are coupled by the set of local chemical reactions in the electron transport complexes of mitochondrial membranes that link proton and electron flows, eventually to ATP production in an ATP synthase Complex 5. The complexes of the electron transport chain function together as one system coupled by both chemical reactions and physical laws.

These local chemical reactions also involve intra-molecular, atomic scale electrical potentials as they appear in the Schrödinger equations that govern their atomic and molecular orbitals as summarized if one wishes in the Hellman-Feynman theorems (Hellmann 1937; Feynman 1939; Di Ventra and Pantelides 2000; Politzer and Murray 2018; Pathak et al. 2023), but that potential and its variation is beyond the resolution of the analysis in this paper. In a higher resolution treatment, biologically significant couplings involving changes in the potential in the Schrödinger equations might emerge. A glimpse of such effects is seen in the multiple chemical guises of 'protons', as positively charged water of with different configurations of chemical bonds with different potentials in the Schrödinger equations of their molecular orbitals.(Yamashita and Voth 2012; Knight and Voth 2011; Knight et al. 2010; Chen et al. 2007)

**Methods advocated here are practical.** The research group of Huaxiong Huang, Xu et al. 2022, have performed actual calculations of coupled currents, without particular difficulties, for one of the complexes (Respiratory Complex 4, the cytochrome c oxidase system) using standard numerical methods of partial differential equations applied to 'multiphysics' systems, where migration and diffusion occur in specialized structures.(Eisenberg 2022c; Xu et al. 2023a, 2023b; Song et al. 2022; Zhu et al. 2021b, 2021a; Zhu et al. 2019; Xu et al. 2018)

The work of Xu et al. 2022 is a practical application of the general ideas in this paper and shows that a respiratory complex can be analyzed by analysis of currents and chemical reactions without explicit discussion of charges or use of the continuity equation (4). Indeed, it is not clear that the respiratory complex can be analyzed without explicit use of the Maxwell and Kirchhoff current law. A charge-based analysis of circuits is not available in the literature of electrical or electronic engineering fundamentally because there are too many charges to deal with. The same is true in mitochondria, or nerve fibers for that matter, as well as circuits, of course. Current laws are always used, in addition to occasional use of the properties of the charges themselves.

**Chemiosmotic Hypothesis, Revisited.** Maxwell's Current Law (2) provides another interpretation (Xu et al. 2022) of the Chemiosmotic Hypothesis of Peter Mitchell (Mitchell and Moyle 1967; Mitchell 1975, 1977; Morelli et al. 2019; Juhaszova et al. 2022; Boyer 1988; Wikstrom et al. 2015). This interpretation allows quantitative prediction of measurable currents that ultimately generate ATP. The ATP is produced by the total (true) Maxwell current, not the protonmotive force.

Note that the units of protonmotive force are not the units of flux or current or ATP production. This difficulty of units in the original chemiosmotic hypothesis is not present when the electro-osmotic approach is used, because the electo-osmotic approach like all circuit analysis is automatically nonequilibrium, with velocity, flows, friction, and power transfer as automatic consequences of the analysis and mathematics. (Eisenberg 2022c; Xu et al. 2023a, 2023b; Song et al. 2022; Zhu et al. 2021b, 2021a; Zhu et al. 2019; Xu et al. 2018)

Using Maxwell's Current Law (2), the chemiosmotic electrical process can now be viewed in the electro-osmotic approach as the flow of *total* (true) current across the electron transport complexes and lipid membrane. (These methods have also been used by (Xu et al. 2023b) to analyze physical systems of the general Butler-Vollmer class.)

Perplexing issues of complexity in electron and proton flow (Morelli et al. 2019; Juhaszova et al. 2022; Boyer 1988; Wikstrom et al. 2015) do not have to be explained by words alone in the electro-osmotic hypothesis, in contrast to the chemiosmotic hypothesis which is notable for its lack of specific quantitative predictions of flows or other nonequilibrium phenomena. In the electro-osmotic approach, flows can be estimated by the Maxwell Current Law (2) just as the perplexing issues of the mechanism of ion current flow through $Na^+$ and $K^+$ selective channels have been analyzed for many years.(Huxley 1972) Imagine how difficult it would be to understand the coupling of $Na^+$ and $K^+$ currents (Hodgkin and Huxley 1952) that produce the propagating action potential if one did not use current laws!

**Respiratory Complex Five and ATP production.** The total current produced by Respiratory Complex 4 generates ATP in the synthase enzyme of Complex 5. It is the total current that generates the ATP, not just the proton flux. An explicit treatment of the coupling of Complex 4 and 5 is a goal as we extend the work of (Xu et al. 2022).

Extended this way, the chemiosmotic hypothesis (Morelli et al. 2019; Juhaszova et al. 2022; Boyer 1988; Wikstrom et al. 2015; Mitchell and Moyle 1967; Mitchell 1975, 1977; Stryer 1995; Mitchell 1961) fits into *and takes advantage of* the general treatment of electricity in the physical sciences (Maxwell 1865b; Maxwell 1865a; Maxwell 1865c; Whittaker 1951; Feynman, Leighton, and Sands 1963b) just as the qualitative understanding of the action potential (Hodgkin 1937b, 1937a) fits into the general treatment of electricity and takes advantage of the cable equation.(Hodgkin and Huxley 1952; Hodgkin and Rushton 1946; Davis and de No 1947; Kelvin 1855, 1856)

In the original chemiosmotic hypothesis (Morelli et al. 2019; Juhaszova et al. 2022; Boyer 1988; Wikstrom et al. 2015; Mitchell and Moyle 1967; Mitchell 1975, 1977; Stryer 1995; Mitchell 1961), conduction currents did not fit so naturally. *Flows of electrons and protons (and currents) were not identified as the currents of physics, as defined in the Maxwell equations of electrodynamics, e.g. eq. (1).*

Flows (and thus currents) of electrons, protons, and ions were hard to specify in traditional models because they vary so much in composition, and in space. Displacement currents did not appear at all, despite their universal existence, according to the Maxwell Ampere equation. The word 'current' was not identified with the currents of classical electrodynamics in the original chemiosmotic hypothesis. A current law was not used to explicitly describe the flow of ions (or electrons or protons) unlike in the treatment of the action potential, done at much the same time, where a current law was used from early on to describe the flow of ions and material displacement current across the cell membrane (Hodgkin and Rushton 1946; Davis and de No 1947), although not from the beginning.(Hodgkin 1937b, 1937a)

**A current law is needed**, in my view, that applies to the complexity of three-dimensional mitochondria where current carriers switch from electrons, to ions, to protons (or hydronium ions (Boyer 1988)). Analysis based on charges and the continuity equation (4) is not practical because all the charges in a mitochondrion are involved and that is a very large number, more or less impossible to compute directly when interactions are included, that are required by Coulomb's law.

**Protonation.** In the chemiosmotic hypothesis, the flows of protons are multifaceted, particularly hard to discern and difficult to deal with because protons might protonate weak bases and interact with acids and bases throughout the mitochondria. The protons would then vary in concentration, changing the conduction current of Kirchhoff's current law. Treatments of total current sidestep the issue of protonation because they deal with the entire true current at once without worrying about its components. The flows of electrons are also multifaceted, hard to deal with as they move through so many systems.

The multifaceted flows of electrons are also a characteristic of current flow in the integrated circuits of our technology, where the properties of electron flow in wires, resistors, transistors, and capacitors are quite different. Current laws successfully bypass such complexities in integrated circuits and are likely to help bypass complexities in mitochondria as well in many cases.

Of course, there are situations where the chemical identity of the charge carriers is important. Then, analysis of total current is not enough, although such analysis remains helpful. Analogous situations arise in the analysis of the action potential, particularly when the validity of voltage clamp studies is being checked by comparisons of current and flux of $Na^+$ ions .(Atwater, Bezanilla, and Rojas 1969, 1970; Bezanilla, Rojas, and Taylor 1970b; Bezanilla, Rojas, and Taylor 1970a; Rojas, Bezanilla, and Taylor 1970)

**Chemiosmotic Hypothesis, revised into an Electro-osmotic approach.** In this revised version of the chemiosmotic theory, the rules are simple. Current is now the total true current of the Maxwell equations, e.g. eq. (1). Total current does not accumulate. The Maxwell Current Law (2) is quantitative, indeed exact. It replaces the rather vague discussion of how complexes in mitochondria interact (found in the chemisosmotic hypothesis (Morelli et al. 2019; Juhaszova et al. 2022; Boyer 1988; Wikstrom et al. 2015; Mitchell and Moyle 1967; Mitchell 1975, 1977; Stryer 1995; Mitchell 1961)) with the precision afforded by nearly two centuries of work on the physics and electrodynamics of current flow.

To the extent that total current describes the flow of energy in the mitochondria, details of the charge movements in conduction currents are unimportant, just as details of the charge movements in the conduction currents are rather unimportant in the circuits of our computers (Eisenberg 2022b) or in the propagation of the actional potential of nerve and muscle, as described below.

**Total current and protein selectivity.** The total current provides the input to the enzymes, transporters, and channels of the complexes of the electron transport chain. The proteins of the complexes can provide the selectivity to choose the form of charge needed as a substrate for the chemical reaction or transport properties of the protein. The current itself need not be carried by a particular charged species in this case just as the identity of the current carrier is unimportant in the conduction of the action potential from place to place, just as the identity of the current carrier in semiconductor circuits is not of primary importance.

In this revised approach, the Maxwell Current Law (2) shows precisely how the total current couples the output of Respiratory Complex 4 to the synthesis of ATP in Respiratory Complex 5 although a full model requires an extension of our previous work (Xu et al. 2022).

The input to the ATPase of Complex 5 is a true current. It cannot accumulate because it is true current as Maxwell defined it. Driven by the protonmotive forces, including the electrical potential of the Maxwell equations, the true current carries protons that help generate ATP while obeying the laws of electrodynamics as it does so, eq. (1)–(5). Those laws guarantee that the electrical potential of the protonmotive forces is coupled to the movement of all charges and displacement currents. Most of the details of the charge movements do not matter. *It is the true total current that provides the energy that generates the ATP, not just the protonmotive force.*

**Ionic Signals in Biology.** The idea of total current is useful in another biological system, the signaling system of nerve and muscle. The total current of the Maxwell Current Law (2) helps define the circuit models familiar in the treatment of the nerve action potential (Hodgkin and Rushton 1946; Davis and de No 1947), and the action potential of cardiac and skeletal muscle (Huxley 1972).

The circuits of nerve and muscle have been ***derived*** by mathematics alone. The derivation starts with the Green's functions of a structural description of complex tissues (Eisenberg Robert 2011; Eisenberg 2018; Eisenberg 2022c) using Taylor expansions (Eisenberg and Johnson 1970; Barcilon, Cole, and Eisenberg 1971; Eisenberg 2022c; Eisenberg 1967) thereby linking the biological treatment of electricity and the physical treatment of electrodynamics. Singular perturbation techniques (Peskoff and Eisenberg 1973; Kevorkian and Cole 1996), p. 218-238, are helpful in applying the Maxwell Current Law (2) to biological systems of some complexity (Barcilon, Cole, and Eisenberg 1971; Eisenberg, Barcilon, and Mathias 1979; Eisenberg 2022c) but it is important to know that all results have been derived by Taylor series as well as perturbation methods, with careful error analysis (Barcilon, personal communication to Bob Eisenberg). The general approach is reviewed in the biological context in (Eisenberg 2022c).

**Approximations inherent in the circuit representation** of nerve and muscle have been extensively examined by many methods. The first approximation involves the cell interior. Experiments (Mobley, Leung, and Eisenberg 1974, 1975; Chandler, Hodgkin, and Meves 1965; Baker, Hodgkin, and Meves 1964) show that the interior of cells can be approximated as resistors. Only the resistance is important. The specific ionic contents of the interior do not matter very much (Eisenberg et al. 2018; Eisenberg 2019c; Chandler, Hodgkin, and Meves 1965; Baker, Hodgkin, and Shaw 1962b, 1962a).

The second approximation in the circuit representation represents the membrane as a capacitance. Experiments show that membranes have well defined capacitances.(Hanai, Haydon, and Taylor 1965b, 1965a; Everitt and Haydon 1968; Fricke 1925b, 1925a; Fricke and Morse 1925) The capacitances arise in lipids that provide pathways for displacement currents $\varepsilon_m \varepsilon_0\, \partial \mathbf{E}/\partial t$ where $\varepsilon_m$ is the effective dielectric constant of the lipid membrane.

The third approximation in the circuit shows that the ion channels and transporters do not change the membrane capacitance under most conditions. They add current to the lipid membrane current (Cole and Curtis 1939) without changing it very much (Bezanilla 2018; Catacuzzeno, Franciolini, et al. 2021; Catacuzzeno, Sforna, et al. 2021) except when gating currents (Bezanilla 2018) are created by specialized structures within membrane proteins that sense voltage.(Catacuzzeno and Franciolini 2022; Bezanilla 2002)

It is historically interesting that the analysis of the action potential depends on the Maxwell Current Law (2) in the form of the cable equation used in 1850 (Kelvin 1855, 1856) to describe the telegraph under the ocean (Gordon 2008), the trans-Atlantic cable. Hodgkin and Huxley used the cable equation (Hodgkin and Rushton 1946; Davis and de No 1947) and Cole's voltage clamp (Huxley 1972) to show that action potentials arose from conduction currents through channels (then called conductances) in the membrane of nerve fibers. The conduction currents cross the membrane inside distinct proteins, called ion channels, which are selective and carry either $Na^+$ or $K^+$ currents flow. The currents are then converted into longitudinal currents in the cytoplasm that spread the action potential down a nerve axon. The axonal currents are carried by whatever ions happen to be in the axoplasm. The chemical identity of the ions that carry the longitudinal current is not important for conduction.(Chandler, Hodgkin, and Meves 1965; Baker, Hodgkin, and Shaw 1962b, 1962a)

Early leaders in biophysics (Hill 1932) thought these conduction currents would be propagated by coupled chemical reactions, but we now know that the channels that conduct those currents are not chemically coupled. They are electrically coupled.(Hodgkin 1937b, 1937a) They are too far apart to interact even through their ionic atmospheres.(Chazalviel 1999)

In fact, only electricity couples the conduction currents of the action potential. The conduction currents are coupled by the global electrical potential (Hodgkin and Huxley 1952) of the cable equation. The electrical coupling creates the propagating action potentials from the otherwise independent conduction currents (Huxley 1972); Fig. 10 and eq. 11 of (Hodgkin, Huxley, and Katz 1952)). The potential change accompanying inward sodium $Na^+$ currents at one location is spread by total longitudinal current—described by the cable equation—to other locations, where the potential change opens sodium channel proteins and produce inward $Na^+$ currents of their own, that in turn propagate further. The type of ions that carry the total longitudinal current are not important.(Chandler, Hodgkin, and Meves 1965; Baker, Hodgkin, and Shaw 1962b, 1962a)

The energy for the action potential comes from gradients of concentration maintained by other systems of membrane proteins called pumps or transporters that use the hydrolysis of ATP as their ultimate source of energy. The gradients of concentration of $Na^+$ power the flows of ions modulated by the channel proteins that are the conduction currents that propagate the action potential, as described above. The ATP hydrolyzed by pumps and transporters is created by the electron transport systems of mitochondria and chloroplasts.

The treatment of the electro-osmotic model allows the chemiosmotic hypothesis to take advantage of the knowledge of current flow in the physical and engineering sciences, particularly its Kirchhoff and Maxwell Current Laws. Knowing the current means knowing an important part of the mechanism of ATP synthesis.

**Practical Implications**. Although the approach and form of equations presented here may suggest radical change in traditional analysis, none is needed in most cases. Many analyses assume a single dielectric constant, use Kirchhoff's current law, and supplement idealized circuits with stray capacitances. Those applications give formulae nearly the same as those here if the material displacement current $(\varepsilon_r - 1)\varepsilon_0\, \partial \mathbf{E}/\partial t$ is added to vacuum displacement current $\varepsilon_0\, \partial \mathbf{E}/\partial t$ of our equations and the appropriate definitions of current are used in classical Kirchhoff laws to avoid double counting.

Biological applications are more dramatically affected. The Maxwell Current Law allows systems not traditionally analyzed by electrodynamics—like the Electron Transport Chains of mitochondria—to be analyzed as circuits. Without the current law, the complexity of mitochondria makes quantitative analysis and predictions of currents nearly impossible. There are just too many charges and interactions to do that. The number of currents is much less than the number of charges, let alone the number of interactions of charges.

## Acknowledgement

Dave Ferry has taught me much of what I know about electronics and displacement currents, and I am most grateful for his help through many years.

It is a daily pleasure to thank Ardyth Eisenberg for all she contributes to my life and work, including to this paper.